\documentclass[showpacs,amsfonts,amsmath,amssymb,aps,pra]{revtex4}
\usepackage{bm}
\usepackage{amsfonts}
\usepackage{graphicx}
\begin{document}
\hbadness=10000

\title{Vortex lines of the electromagnetic field}
\author{Iwo Bialynicki-Birula}
 \email{birula@cft.edu.pl}
\affiliation{Center for Theoretical Physics, Polish Academy of Sciences,\\
Lotnikow 32/46, 02-668 Warsaw, Poland\\ and Institute of Theoretical Physics,
Warsaw University}
\author{Zofia Bialynicka-Birula}
\affiliation{Institute of Physics, Polish Academy of Sciences and College of Science,\\
Al. Lotnik\'ow 32/46, 02-668 Warsaw, Poland}
\begin{abstract}
Relativistic definition of the phase of the electromagnetic field, involving
two Lorentz invariants, based on the Riemann-Silberstein vector is adopted to
extend our previous study [I. Bialynicki-Birula, Z. Bialynicka-Birula and C.
{\'S}liwa, Phys. Rev. A {\bf 61}, 032110 (2000)] of the motion of vortex lines
embedded in the solutions of wave equations from Schr\"odinger wave mechanics
to Maxwell theory. It is shown that time evolution of vortex lines has
universal features; in Maxwell theory it is very similar to that in
Schr\"odinger wave mechanics. Connection with some early work on
geometrodynamics is established. Simple examples of solutions of Maxwell
equations with embedded vortex lines are given. Vortex lines in
Laguerre-Gaussian beams are treated in some detail.
\end{abstract}
 \maketitle

\section{Introduction}

The physical significance of the singularities of the phase of quantum
mechanical wave functions has been recognized by Dirac in his work on magnetic
monopoles \cite{dirac}. The hydrodynamic formulation of the Schr\"odinger
theory discovered by Madelung \cite{madelung} provided a vivid interpretation
of the lines in space where the phase is singular. These are simply the vortex
lines in the flow of the probability fluid. The velocity field ${\bm v({\bm r},
t)}$ of this fluid, defined in terms of the probability current ${\bm j}$, is
equal to the gradient of the phase $S$ of the wave function
$\psi=R\exp(iS/\hbar)$,
\begin{eqnarray}\label{velocity0}
{\bm v} = \frac{{\bm j}}{\rho} =
\frac{\hbar}{2mi}\frac{\psi^*{\bm\nabla}\psi-{\bm\nabla}\psi^*\psi}
{\vert\psi\vert^2} = \frac{{\bm\nabla}S}{m}.
\end{eqnarray}
Therefore, the flow is strictly irrotational in the bulk; vorticity may live
only on the lines of singularities of the phase. Regular wave functions may
have a singular phase only where the wave function vanishes, i.e. where
$\Re\psi = 0$ and $\Im\psi = 0$. These two equations define two surfaces in
space whose intersection determines the position of vortex lines. However, the
vanishing of the wave function is the necessary but not the sufficient
condition for the existence of vortex lines. They exist only if the circulation
around the line where the wave function vanishes is different from zero. The
univaluedness of the wave function requires the quantization of the circulation
\begin{eqnarray}\label{condition}
\oint{\bm dl}\!\cdot\!{\bm v} = 2\pi n\hbar/m.
\end{eqnarray}
The importance of this condition in the hydrodynamic formulation of wave
mechanics has been elucidated for the first time by Takabayasi
\cite{takabayasi}. If Eq.~(\ref{condition}) holds for every closed contour,
we may recover the phase $S$ (modulo $2\pi\hbar$) from ${\bm v}$ up to a
global, constant phase with the help of the formula
\begin{eqnarray}\label{phase0}
S({\bm r}) = m\int_{{\bm r}_0}^{{\bm r}}{\bm dl}\!\cdot\!{\bm v}.
\end{eqnarray}
Early studies of vortex lines were restricted to wave mechanics but Nye and
Berry \cite{nye_berry,berry,nye_book,berrySPIE} have shown that phase
singularities or wavefront dislocations play an important role not only in wave
mechanics but in all wave theories. A general review of phase singularities in
wave fields has been recently given by Dennis \cite{dennis,dennis1}. There is a
substantial overlap of concepts (but not of the results) between our work and
the works of Berry, Nye and Dennis. While they concentrate mostly on the
stationary vortex lines that are found in monochromatic fields, we emphasize
the time evolution.

More recently, the study of phase singularities and vortices in optics has
evolved into a separate area of research, both theoretical and experimental,
called singular optics. A recent review of this field is given in
Ref.~\cite{sos_vas}.

In order to find a natural generalization of Eq.~(\ref{velocity0}), we need a
replacement for the wave function $\psi$ in electromagnetism. A suitable object
appears in the complex form of Maxwell equations known already to Riemann
\cite{riem} and investigated more closely by Silberstein \cite{silber} at the
beginning of the last century. In this formulation the electric and magnetic
field vectors are replaced by a single complex vector ${\bm F}$ that we
proposed to call the Riemann-Silberstein (RS) vector \cite{ibbapp,ibbwf}
\begin{eqnarray}
 {\bm F}= ({\bm E}+i{\bm B})/\sqrt{2}.\label{rs}
\end{eqnarray}
Maxwell equations in free space written in terms of ${\bm F}$ read ($c = 1$)
\begin{subequations}\label{maxwell}
\begin{eqnarray}
i\partial_t{\bm F} &=& \nabla\times{\bm F},\label{maxwell1}\\
\nabla\cdot{\bm F} &=& 0.\label{maxwell2}
\end{eqnarray}
\end{subequations}
The analogy between Eq.~(\ref{maxwell1}) and the Schr\"odinger wave equation
is so close that one is lead to treat ${\bm F}$ as the photon wave function
\cite{ibbwf} and apply similar methods to analyze the vortex lines and their
motion as we have done in Refs.~\cite{bbs,bmrs} in nonrelativistic wave
mechanics. There is, however, an important difference that requires an
extension of our previous methods: the RS vector has three components instead
of one. Thus, there are three independent phases $\varphi_1$, $\varphi_2$,
$\varphi_3$
--- one for each component and it is not clear which combination of
these phases should be treated as an overall phase of the electromagnetic
field.

In the case of the the Schr\"odinger wave function, the information about the
phase $S$ of the wave function is stored in the velocity field ${\bm
v}={\bm\nabla}S/m$. Hence, one may try to find the proper definition of the
phase of the electromagnetic field by introducing first the counterpart of
Eq.~(\ref{velocity0}) and then use the velocity field to reconstruct the
phase. The natural generalization of the definition (\ref{velocity0}) is (in
dimensionless form)
\begin{eqnarray}\label{velocity_em}
{\bm v} = \frac{1}{2i}\frac{\sum_k\left(F_k^*{\bm\nabla}F_k-
({\bm\nabla}F_k^*)F_k\right)} {\sum_k F_k^*F_k}.
\end{eqnarray}
However, as has been noticed already by Takabayasi in his study of the
hydrodynamic formulation of wave mechanics of spinning particles
\cite{takabayasi1}, this generalization does not work. For a multicomponent
field the velocity defined in this way cannot be used to reconstruct the phase
because, in general, ${\bm\nabla}\times{\bm v}$ does not vanish. Even though
one can still give a hydrodynamic interpretation of Maxwell theory based on the
formula (\ref{velocity0}), the simplicity of the scalar case is completely lost
\cite{ibb2}.

In the present paper, the phase of the electromagnetic field and the vortex
lines associated with this phase are defined in terms of the square of the
Riemann-Silberstein vector. Since ${\bm F}^2$ is a sum of two electromagnetic
invariants, the structure of phase singularities associated with ${\bm F}^2$ is
relativistically invariant. This definition of the phase turns out to be
equivalent (provided ${\bf F}$ obeys Maxwell equations) to the one used in the
classic papers on geometrodynamics \cite{rainich,mw,witten}.

Despite the fact that ${\bm F}^2$ does not obey any simple wave equation, the
time evolution of the vortices exhibits all the typical features found before
by us for the Schr\"odinger equation. During the time evolution governed by
Maxwell equations vortex lines are created and annihilated at a point or in
pairs and undergo vortex reconnections.

\section{Geometrodynamics and the phase of the electromagnetic field}

In nonrelativistic wave mechanics the phase of the wave function can be
obtained from its modulus provided we also assume that the wave function obeys
the Schr\"odinger equation. As a matter of fact it was shown by E. Feenberg
\cite{kemble} that to determine the phase from the modulus it is sufficient
that the wave function obeys {\em some} wave equation that leads to
conservation of the probability, i.e. to continuity equation. A similar
reasoning applied to the electromagnetic field also enables one to determine
the (properly defined) phase of this field. This discovery has been made by
Rainich \cite{rainich} in connection with the problem of the reconstruction of
the electromagnetic field from purely geometric quantities in general
relativity. Independently, although much later, this problem was solved by
Wheeler and coworkers \cite{mw,witten,mtw,pr} in the context of
geometrodynamics.

Very briefly, the reconstruction of the electromagnetic field from geometry
may be described as follows. The Einstein equations
\begin{eqnarray}\label{einstein}
R_{\mu\nu} - g_{\mu\nu}R/2 = \kappa T_{\mu\nu}
\end{eqnarray}
enable one to determine the energy-momentum tensor $T_{\mu\nu}$ of the
electromagnetic field from the Einstein tensor $R_{\mu\nu} - g_{\mu\nu}R/2$
that is made of the metric tensor and its derivatives. However, the knowledge
of the energy-momentum tensor alone is not sufficient to determine completely
the electromagnetic field. This is best seen from the formulas for the
components of this tensor expressed in terms of the RS vector:
\begin{subequations}\label{en_mom}
\begin{eqnarray}
T_{00} &=& {\bm F}^*\cdot{\bm F},\label{en_mom1}\\
T_{0i} &=& \epsilon_{ijk}F^*_jF_k/i,\label{en_mom2}\\
T_{ij} &=& -F^*_i F_j - F^*_j F_i + \delta_{ij}{\bm F}^*\!\cdot\!{\bm
F}.\label{en_mom3}
\end{eqnarray}
\end{subequations}
All components of the energy-momentum tensor are invariant under the common
change of the phase of all three components of the RS vector --- the duality
transformation
\begin{subequations}\label{dual_rot}
\begin{eqnarray}
 {\bm E}' = {\bm E}\cos\varphi-{\bm B}\sin\varphi,\\
 {\bm B}' = {\bm E}\sin\varphi+{\bm B}\cos\varphi.
\end{eqnarray}
\end{subequations}
Therefore, the overall phase cannot be determined from the energy-momentum
tensor. Note, that in contrast to the situation in quantum mechanics, even the
global, constant phase of ${\bm F}$ has a direct physical meaning. It controls
the relative contribution to the energy-momentum tensor from the electric and
the magnetic parts. The duality rotations (\ref{dual_rot}) with a constant
value of $\varphi$ leave the free Maxwell equations unchanged. However, a phase
varying in space and/or time would modify the Maxwell equations. That is the
reason why the Rainich construction works. Namely, he has shown that if one
assumes that the electromagnetic field obeys Maxwell equations, the phase of
the field may be extracted from $T_{\mu\nu}$. For this purpose he introduced
the following four-vector built from the components of the energy-momentum
tensor and its derivatives
\begin{eqnarray}\label{velocity2}
w^\mu = -\frac{\epsilon^{\mu\nu\lambda\rho}T_{\nu\kappa}\partial_\lambda
T^{\kappa}_{\;\;\rho}}{T_\sigma^{\;\tau}T_\tau^{\;\sigma}}
\end{eqnarray}
and used the line integral of $w^\mu$ to reconstruct the phase.

Our proposal, how to define the phase of the electromagnetic field is much
simpler and yet it turns out to be completely equivalent to the definition
given by Rainich. We shall define the phase of the electromagnetic field
$\varphi(x)$ as half of the phase of the square of the RS vector
\begin{eqnarray}\label{phase1}
{\bm F}^2(x) = e^{2i\varphi(x)}\vert{\bm F}^2(x)\vert.
\end{eqnarray}
In full analogy with Eq.~(\ref{velocity0}) of nonrelativistic wave mechanics,
we define a ``velocity'' four-vector $u^\mu$ as
\begin{eqnarray}\label{velocity1}
u_\mu = \frac{({\bm F}^2)^*\partial_\mu{\bm F}^2 - {\bm F}^2\partial_\mu({\bm
F}^2)^*}{4i\vert{\bm F}^2\vert^2} = \partial_\mu\varphi(x).
\end{eqnarray}
Since ${\bm F}^2$ is a complex sum of two electromagnetic invariants
\begin{eqnarray}\label{square}
 {\bm F}^2 = {\cal S} + i{\cal P} = \frac{1}{2}({\bm
 E}^2 - {\bm B}^2) + i{\bm E}\!\cdot\!{\bm B},
\end{eqnarray}
$u_\mu$ is a true relativistic four-vector
\begin{eqnarray}\label{velocity3}
u_\mu =  \frac{{\cal S}\partial_\mu{\cal P}-{\cal P}\partial_\mu{\cal
S}}{2({\cal S}^2+{\cal P}^2)}.
\end{eqnarray}
This vector has the same denominator (up to a factor of 2 that scales both the
numerator and the denominator) as the vector $w^\mu$ defined by
Eq.~(\ref{velocity2}) since $T_\sigma^{\;\tau}T_\tau^{\;\sigma} = 4({\cal S}^2
+ {\cal P}^2)$. However, in general, the numerators of vectors $w_\mu$ and
$u_\mu$ are different. They do become equal when the electromagnetic field
obeys the Maxwell equations. The proof is straightforward but rather tedious
and will not be presented here.

In our formulation, the square of the RS vector plays the role of the wave
function $\psi$. Vortex lines are to be found at the intersection of the ${\cal
S} = 0$ and ${\cal P} = 0$ surfaces. As in the case of the Schr\"odinger wave
function, at all points where ${\bm F}^2$ does not vanish, the vector $u_\mu$
is {\em by construction} a pure gradient
\begin{eqnarray}\label{gradient}
u_\mu(x) = \partial_\mu\varphi(x).
\end{eqnarray}
Therefore, one may recover the phase of ${\bm F}$ by the following line
integral
\begin{eqnarray}\label{line}
\varphi(x) = \int_{\bm x_0}^{\bm x}\!d{\xi}^\mu u_\mu({\xi}).
\end{eqnarray}
Since the RS vector is univalued, the phases obtained by choosing different
paths connecting the points $x$ and $x_0$ may differ only by a multiple of
$2\pi$. In other words, the vorticity associated with $u_\mu$ (or with $w_\mu$
in the Rainich construction) must be quantized
\begin{eqnarray}\label{circle}
\oint\!d{\xi}^\mu u_\mu({\xi}) = 2\pi n.
\end{eqnarray}
The phase defined by Eq.~(\ref{line}) is determined up to a global phase
$\varphi_0$: the value of $\varphi(x)$ at the lower limit $x_0$ of the
integral. The value of $\varphi_0$ cannot be obtained from the energy-momentum
tensor.

Under duality rotations (\ref{dual_rot}) when $\varphi$ varies from 0 to
$2\pi$, the vector ${\bm E}'$ at each spacetime point draws an ellipse in the
${\bm E}-{\bm B}$ plane. The same ellipse is drawn by the vector ${\bm B}'$.
These ellipses become circles on each vortex line since then the vectors ${\bm
E}$ and ${\bm B}$ are orthogonal and of equal length. This property lead Berry
and Dennis \cite{dennis} to name the vortex lines associated with the square of
a complex vector field the C (circle) lines in their general classification
scheme of phase singularities.

The denominator in Eq.~(\ref{velocity3}) may be also expressed in the form
\begin{eqnarray}\label{four-vector}
{\cal S}^2+{\cal P}^2 = \left(\frac{{\bm E}^2+{\bm B}^2}{2}\right)^2
 - ({\bm E}\times{\bm B})^2.
\end{eqnarray}
Therefore, the vanishing of ${\cal S}^2+{\cal P}^2$ at a point also means that
the electromagnetic field at this point is pure radiation: the energy density
and the Poynting vector form a null four-vector. One may say that on vortex
lines the energy of the electromagnetic field moves locally with the speed of
light. We would like to emphasize that the velocity of the energy flow of the
electromagnetic field is not correlated with the vector $u_\mu$. Even the
geometric properties of the Poynting vector and the space part of $u_\mu$ are
different. Since ${\cal S}$ is a scalar and ${\cal P}$ is a pseudoscalar, the
vector $u_\mu$ is a pseudovector. In the simplest case of a constant
electromagnetic field the Poynting vector is ${\bm E}\times{\bm B}$, while the
vector $u_\mu$ vanishes identically. There does not seem to exist a physical
quantity whose flow can be identified with $u_\mu$. In this respect the
situation is quite different from nonrelativistic wave mechanics where the
gradient of the phase determines the velocity of the probability flow.

\section{Simple examples of vortex lines}

The analogy between the phase of wave function and the phase of the
electromagnetic field is not exact. Unlike the Schr\"odinger wave function, the
electromagnetic field does not have to vanish {\em identically} along the lines
where the phase is singular. It is only necessary that the field is null i.e.
the two invariants ${\cal S}$ and ${\cal P}$ vanish. Still, we believe that the
lines along which the field is null deserve the name of vortex lines.

The time evolution of the vortex lines embedded in the solutions of the Maxwell
equations is quite similar to the evolution of such lines embedded in the
solutions of the Schr\"odinger equation. The simplest examples of solutions
with vortex lines can again be found among the polynomial functions. Such
functions may be viewed as long wavelength expansions and were found to be very
useful in the study of vortex solutions of the Schr\"odinger equation
\cite{bbs,bmrs} and the Helmholtz equation \cite{nye_berry,berry_dennis}.
Alternatively, these polynomial solutions may be viewed as local approximations
to the full solution, valid close to the vortex lines under study. In this case
one may imagine that in the exact solution the polynomial is multiplied by some
slowly varying envelope that makes the full solution localized. We shall give
at the end of this Section an example of such a solution.

\begin{figure}[!]{
\includegraphics[width=8cm]{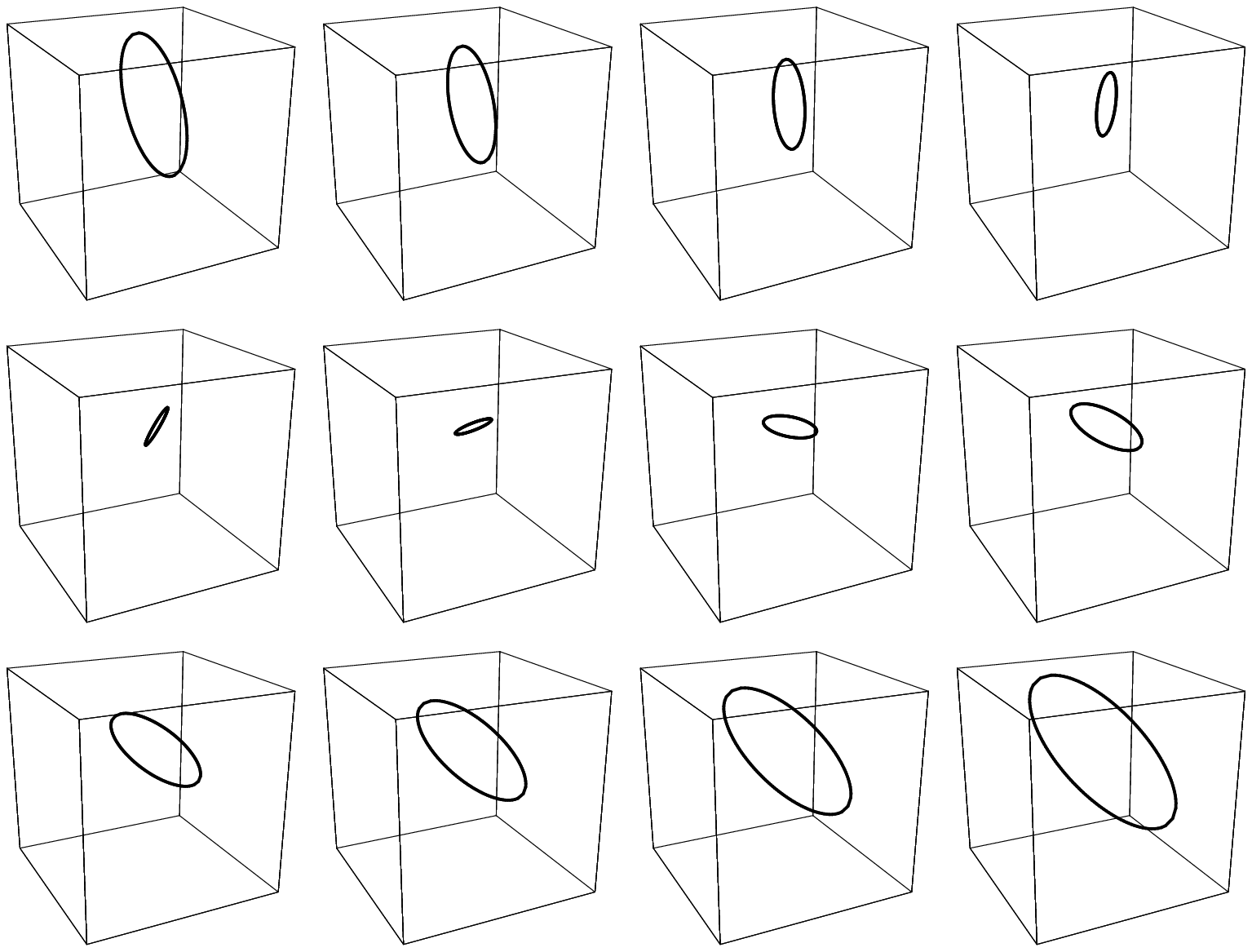}
\caption{Time evolution of a vortex line in the form of a rotating and
expanding ring. All figures in this paper were produced with the use of
Mathematica \cite{wolfram}.}\label{fig:ring}}
\end{figure}

As an illustration of a typical behavior of electromagnetic vortex lines, we
present very simple examples of the electromagnetic field. The following four
fields satisfy the Maxwell equations and possess the vortex structures very
similar to those found in Schr\"odinger wave mechanics \cite{bbs,bmrs}
\begin{widetext}
\begin{subequations}\label{four_solns}
\begin{eqnarray}
{\bm F}^{(a)} &=& \{y + it, z - a + i(a + t), x + it\},\\
{\bm F}^{(b)} &=& \{y + t, a - i(z + a + t), x + it\},\\
{\bm F}^{(c)} &=& \{2x + y - a + i(z + y + t), z - y + t + i(y - a),-t + it\},\\
{\bm F}^{(d)} &=& \{z^2 + t^2 - iat, a^2 - i(2zt + a^2 + ax),a(y - t)\},
\end{eqnarray}
\end{subequations}
\end{widetext} where $a$ is a parameter that sets the scale for the vortex
configuration. In the first three cases the electromagnetic fields are linear
functions of the coordinates and in the last case the field is quadratic. In
the first case, the two invariants are
\begin{subequations}\label{two_inv}
\begin{eqnarray}
{\cal S} &=& x^2 + y^2 + (z - a)^2 - a^2 -2at - 3t^2,\\
{\cal P} &=& 2az + 2t(x + y + z - a) - 2a^2.
\end{eqnarray}
\end{subequations}
The equations ${\cal S}=0$ and ${\cal P}=0$ describe a sphere centered at the
point $(0,0,a)$ with the time dependent radius $\sqrt{a^2 + 2at + 3t^2}$ and a
moving plane, respectively. The intersection of these two surfaces is a moving
ring shown in Fig.~\ref{fig:ring}. The radius of the sphere decreases for
negative values of $t$ until $t = -a/3$ and then starts increasing. The rate of
change of the radius exceeds (by a factor of $\sqrt{3}$) the speed of light
showing once again that various characteristic features of relativistic fields
(like their zeros or maxima) may travel with superluminal speeds without
violating causality. In this simple example, no change of the topology of
vortex line takes place. However, in the three remaining cases the topology
changes according to the same universal patterns as those found in
Schr\"odinger wave mechanics. This universal behaviour of vortex lines is
reminiscent of the catastrophe theory \cite{nye1,berry1}.

\begin{figure}[!]{
\includegraphics[width=8cm]{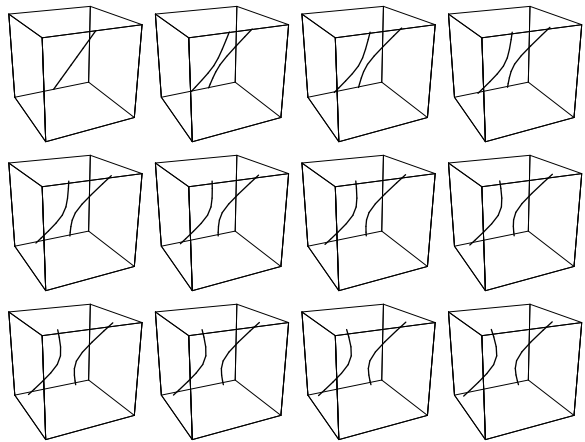}
\caption{Time evolution of two vortex lines that at $t=a$ suddenly appear as a
straight line and then separate and fly away. The same solution for negative
times would show a reversed process: the convergence of two vortex line and
their annihilation at $t=-a$.} \label{fig:lines}}
\end{figure}

\begin{figure}[!]{
\includegraphics[width=8cm]{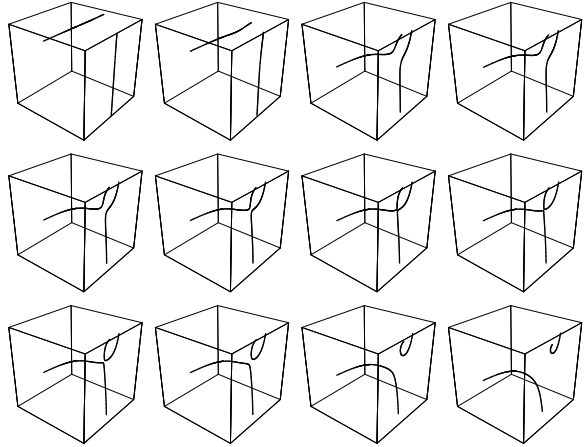}
\caption{Time evolution of two vortex lines that at $t=0$ are mutually
perpendicular and nonintersecting. At the time $t=a(\sqrt{2}-1)^{1/2}$ the
vortex lines cross and undergo a reconnection.} \label{fig:reconnect}}
\end{figure}

\begin{figure}[!]{
\includegraphics[width=8cm]{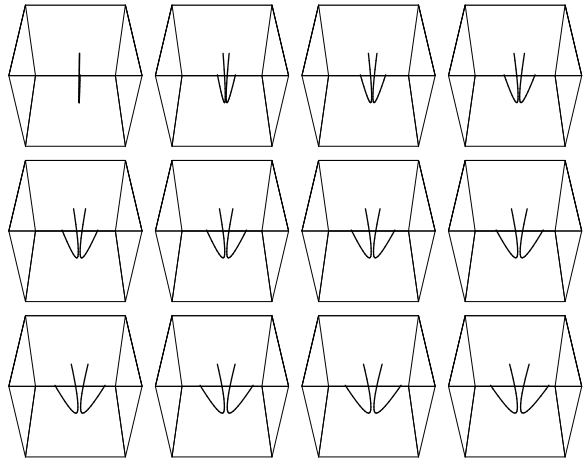}
\caption{Time evolution of two vortex lines that at $t=a$ suddenly appear along
a parabola in the $y=a$ plane (shown as a line in the first frame) that first
opens up very fast into two symmetric wings that later slowly separate.}
\label{fig:butterfly}}
\end{figure}

The graphical representation of the motion of the vortex lines in all four
cases is straightforward since the equations ${\cal S} = 0$ and ${\cal P} = 0$
can be solved analytically giving $x$ and $y$ for each value of $t$ as
parametric functions of $z$. In each case there are two branches that differ by
the sign of the square root.
\begin{widetext}
\begin{subequations}\label{four_curves}
\begin{eqnarray}
x(z,t) &=& \left((a -z)(a +t) \pm\sqrt{a^2 + 2at + 3t^2}
        \sqrt{2t^2 -(a - z)^2}\right)/(2t),\nonumber\\
y(z,t) &=& \left((a + t)(a - z) \mp\sqrt{a^2 + 2at + 3t^2}
        \sqrt{2t^2 -(a - z)^2}\right)/(2t),\label{ring}\\
 ~\nonumber\\
x(z,t) &=& a(a + z)/t - a,\nonumber\\
y(z,t) &=&  -t \pm \sqrt{t^2 - a^2}
     \sqrt{a^2 - 2at + 2t^2 + 2az - 2zt + z^2}/t,\label{lines}\\
 ~\nonumber\\
x(z,t) &=& t + \left(t^2
    \pm\sqrt{t^4 - 8t^2(z - a)z - 16(z - a)^2z^2}\right)/(4(z-a)),
    \nonumber\\
y(z,t) &=& (a - z)/2 + \left(t^2
    \mp\sqrt{t^4 - 8\,t^2(z - a)z - 16(z - a)^2z^2}\right)/(8z),\label{reconnect}\\
 ~\nonumber\\
x(z,t) &=& t \pm\sqrt{t^2 - a^2}
   \sqrt{a^4 + ( t^2 + z^2)^2}/a^2,\nonumber\\
y(z,t) &=& -a - \left(t^3+ zt(2a + z)\right)/a^2.\label{butterfly}
\end{eqnarray}
\end{subequations}
\end{widetext}
The plots of the functions (b) and (d) show vortex creations and annihilations
(Fig.~\ref{fig:lines} and Fig.~\ref{fig:butterfly}) and for the functions (c)
one obtains vortex reconnections (Fig.~\ref{fig:reconnect}). Vortex
annihilations occur at $t = -a$ and vortex creations occur at $t = a$. Note
that according to the formulas (\ref{lines}) and (\ref{butterfly}), at these
moments the vortex velocity $(dx/dt,dy/dt)$ becomes infinite.

\begin{figure}[!]{
\includegraphics[width=8cm]{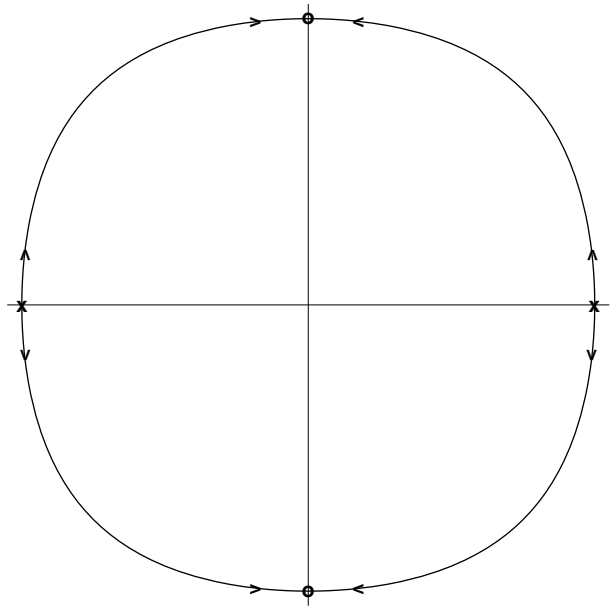}
\caption{Time evolution in the $xy$ plane of two pairs of straight vortex lines
parallel to the $z$ axis. The evolution is indicated by the arrows. Pairs of
vortex lines are created at the points denoted by crosses and annihilated at
the points denoted by circles.} \label{fig:fourvort}}
\end{figure}

It is also possible to construct localized, finite energy solutions of Maxwell
equation with vortices. We shall give just one simple example of such a
solution constructed from the following localized solution of the wave equation
\begin{eqnarray}\label{loc_soln}
 {\bm Z}({\bf r}, t) = \{y,x,-t-i\tau\}((t + i\tau)^2 - {\bf r}^2)^{-2}.
\end{eqnarray}
With each vector solution of the wave equation one may associate a solution of
Maxwell equations treating the solution of the wave equation as a complex
counterpart of the Hertz potential. Namely, one may check the RS vector ${\bm
F}({\bf r}, t)$ constructed according to the following prescription \cite{bb}
\begin{eqnarray}\label{hertz}
 {\bf F}({\bf r}, t) = \nabla\times \left[i\partial_t{\bf Z}({\bf r}, t)
 + \nabla\times{\bf Z}({\bf r}, t)\right]
\end{eqnarray}
indeed satisfies the Maxwell equations. The square of the vector ${\bf F}$ has
the form
\begin{eqnarray}\label{square1}
 32\frac{2(t^2 + 2x^2 + 2y^2 -\tau^2)
  + i(3x^2 - 3y^2 +4t\tau)}{((t + i\tau)^2 - {\bf r}^2)^{6}}.
\end{eqnarray}
Since the numerator does not contain the variable $z$, the vortex lines
embedded in this localized solution are straight lines parallel to the $z$
axis. Two pairs of such lines are created at $t = -\tau/3$ at the points $(\pm
2\tau/3,0)$ in the $xy$ plane. The four vortex lines move
(Fig.~\ref{fig:fourvort}) until they annihilate in pairs at $t = \tau/3$ at the
points $(0,\pm 2\tau/3)$. The speed of each vortex line at the moment of
creation and annihilation is {\em infinite}, showing very vividly that also for
localized solutions of Maxwell equations the motion of vortex lines may be
superluminal without any limitations. Arbitrarily high speed of vortex lines
associated with solutions of the relativistic scalar wave equation has already
been noted in Refs.~\cite{nye_berry,bbs}.

\begin{figure}[!]{
\includegraphics[width=8cm]{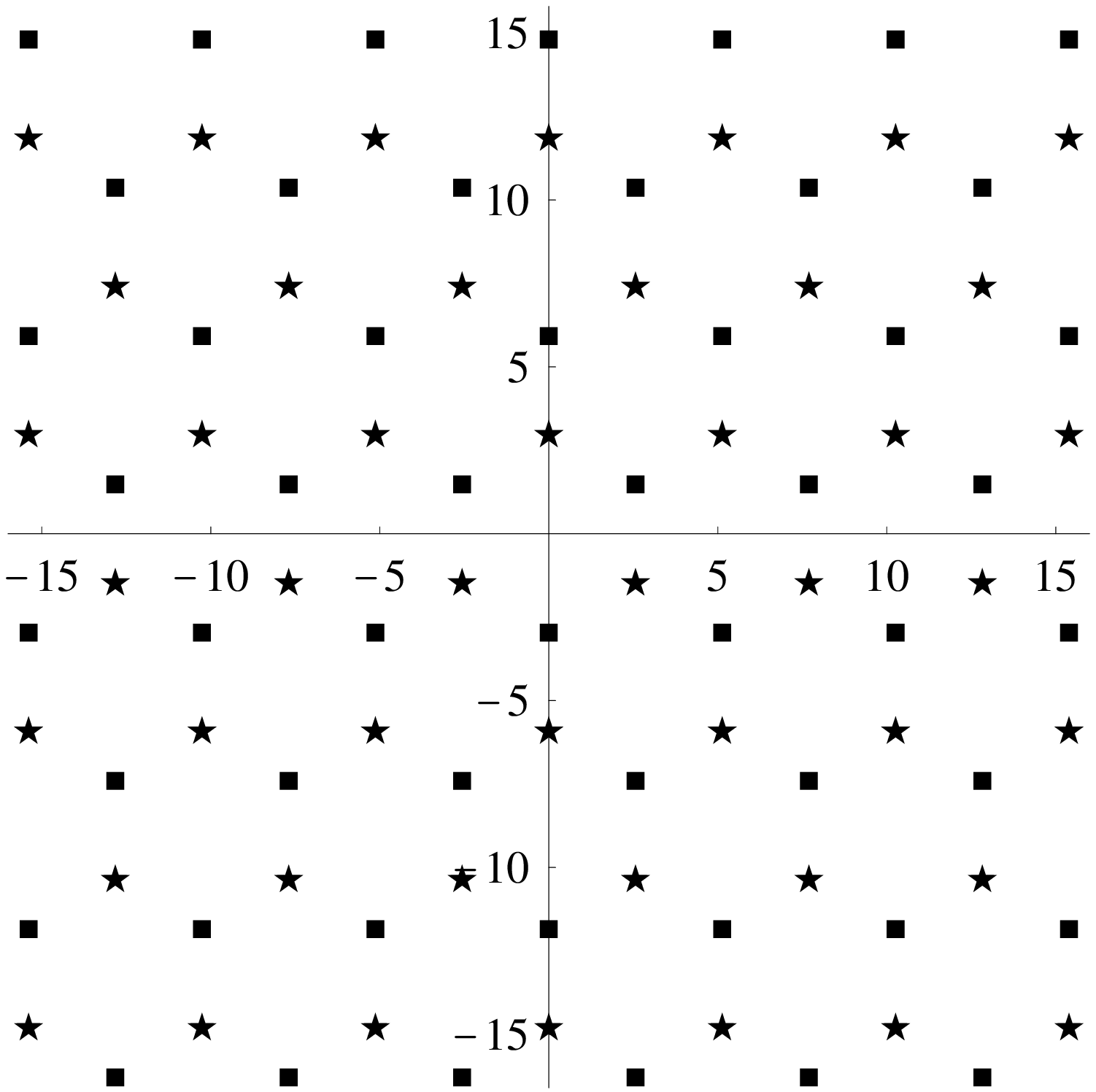}
\caption{Positions of vortex lines in the $xy$ plane (measured in units of the
inverse wave vector) for the three plane waves moving in orthogonal directions.
Points marked with different symbols correspond to vortices with opposite
circulation.}\label{fig:grid}}
\end{figure}

\section{Vortex lines in superpositions of plane waves and in Gaussian beams}

Solutions of Maxwell equations exhibiting vortex structures may also be
obtained with the use of standard building blocks --- the monochromatic plane
waves. A single plane wave is described by a null field since both invariants
vanish. Therefore, the velocity (\ref{velocity3}) vanishes --- a single plane
wave has no vortex structure. Also, the sum of two plane waves does not have
any vortex structure; even though it has a nonvanishing velocity field.
However, for three plane waves we may have various kinds of vortex structures.
As an example, we choose three circularly polarized monochromatic waves of the
same frequency, handedness, and amplitude, moving in three mutually orthogonal
directions. The RS vector in this case (up to a constant amplitude) has the
form
\begin{widetext}
\begin{eqnarray}\label{3waves}
{\bf F}({\bm r}, t) = ({\hat{\bm m}} + i{\hat{\bm n}})e^{-i(t - {\hat{\bm
l}}\cdot{\bm r})} + ({\hat{\bm n}} + i{\hat{\bm l}})e^{-i(t - {\hat{\bm
m}}\cdot{\bm r})} + ({\hat{\bm l}} + i{\hat{\bm m}})e^{-i(t - {\hat{\bm
n}}\cdot{\bm r})},
\end{eqnarray}
\end{widetext}
where ${\hat{\bm l}},{\hat{\bm m}}$, and ${\hat{\bm n}}$ are three orthogonal
unit vectors, the coordinates are measured in units of the inverse wave vector
and time is measured in units of inverse angular frequency. The square of this
vector vanishes at the points satisfying the equation
\begin{eqnarray}\label{3waves_eq}
e^{i({\hat{\bm l}}+{\hat{\bm m}})\cdot{\bm r}} + e^{i({\hat{\bm m}}+{\hat{\bm
n}})\cdot{\bm r}} + e^{i({\hat{\bm n}}+{\hat{\bm l}})\cdot{\bm r}} = 0.
\end{eqnarray}
It is convenient to chose the coordinate system in such a way that the three
basis vectors have the form
\begin{eqnarray}\label{vectors}
{\hat{\bm l}}=\left(\begin{array}{c} \frac{1}{\sqrt{6}}\\
-\frac{1}{\sqrt{2}}\\
\frac{1}{\sqrt{3}}\end{array}\right),\;\;
{\hat{\bm m}}=\left(\begin{array}{c} \frac{1}{\sqrt{6}}\\
\frac{1}{\sqrt{2}}\\
\frac{1}{\sqrt{3}}\end{array}\right),\;\;
{\hat{\bm n}}=\left(\begin{array}{c} -\sqrt{\frac{2}{3}}\\
0\\
\frac{1}{\sqrt{3}}\end{array}\right),
\end{eqnarray}
because then all the vortex lines are parallel to the $z$ axis. The position of
the vortex lines in the $xy$ plane is determined by Eq.~(\ref{3waves_eq}). For
the choice (\ref{vectors}) of unit vectors this equation has the form (apart
from an overall phase-factor $\exp\left(i(\sqrt{2}\,x+2\,z)/\sqrt{3}\right)$)
\begin{eqnarray}\label{positions}
1 + e^{-i(\sqrt{3}x + y)/\sqrt{2}} + e^{-i(\sqrt{3}x - y)/\sqrt{2}} = 0.
\end{eqnarray}
The solutions of this equation are
\begin{eqnarray}\label{solutions}
x^\pm_{mn} = \pi\sqrt{\frac{2}{3}}\;(m + n),\;\;y^\pm_{mn} =
\pi\sqrt{2}\;(\pm\frac{2}{3} + m - n),\;\;
\end{eqnarray}
where $m$ and $n$ are arbitrary integers. The lattice of vortex lines is shown
in Fig.~\ref{fig:grid}. This example shows that vortex lines associated with
the phase of the RS vector do not necessarily move; they can also be
stationary.

When one of the polarizations of the three waves, say the last one in
Eq.~(\ref{3waves}), is opposite, the position of vortex lines is determined by
a time-dependent equation
\begin{eqnarray}\label{3waves_eq1}
e^{i({\hat{\bm l}}+{\hat{\bm m}})\cdot{\bm r} - 2it} + e^{i({\hat{\bm
m}}-{\hat{\bm n}})\cdot{\bm r}} + e^{i({\hat{\bm l}}-{\hat{\bm n}})\cdot{\bm
r}} = 0.
\end{eqnarray}
In this case it is convenient to choose the orthonormal unit vectors in the
form
\begin{eqnarray}\label{vectors1}
{\hat{\bm l}}=\left(\begin{array}{c} \frac{1}{\sqrt{6}}\\
\frac{1}{\sqrt{2}}\\
\frac{1}{\sqrt{3}}\end{array}\right),\;\;
{\hat{\bm m}}=\left(\begin{array}{c} \frac{1}{\sqrt{6}}\\
-\frac{1}{\sqrt{2}}\\
\frac{1}{\sqrt{3}}\end{array}\right),\;\;
{\hat{\bm n}}=\left(\begin{array}{c} \sqrt{\frac{2}{3}}\\
0\\
-\frac{1}{\sqrt{3}}\end{array}\right).
\end{eqnarray}
The position of vortex lines in the $xy$ plane is determined by the equation
\begin{eqnarray}\label{positions1}
1 + e^{-i(\sqrt{3}x + y)/\sqrt{2}+2it}
 + e^{-i(\sqrt{3}x - y)/\sqrt{2}+2it} = 0.
\end{eqnarray}
Thus, in this case the lattice of vortex lines is not stationary but it is
moving as a whole with the speed of $\sqrt{8/3}c$ in the $x$ direction.

The most interesting case, of course, is a superposition not of a few but of a
continuum of plane waves, forming a collimated beam. We shall concentrate on
the Laguerre-Gaussian beams, in view of their applicability to realistic
situations (cf., for example \cite{wright}). We use the representation of these
beams in the vector theory as in Refs.\cite{absw,allen0,allen,sps} but we
combine the electric and the magnetic field vectors into the complex RS vector
(\ref{rs}). This vector for Laguerre-Gaussian beams of circular polarization
can be written in the form
\begin{eqnarray}
{\bm F}(x,y,z,t) = e^{-i(\omega t - k z)} \left(k u, i k u,i(\partial_x u + i
\partial_y u)\right).\label{LG}
\end{eqnarray}
The square of the this vector is equal to
\begin{eqnarray}\label{squareLG}
({\bm F}(x,y,z,t))^2 =
 -e^{-2i(\omega t - k z)} ((\partial_x +i\partial_y)u)^2.
\end{eqnarray}
Note, that the vector ${\bm F}$ given by Eq.~(\ref{LG}) is not just the
analytic signal but the full RS vector as defined by Eq.~(\ref{rs}) whose real
part is the electric field and the imaginary part is the magnetic induction.
The slowly varying complex envelope function $u=u(x,y,z)$ is an arbitrary
linear superposition of the functions $u_{n m}(\rho,\phi,z)$ defined as (we use
the notation of Ref.~\cite{sps})
\begin{widetext}
\begin{eqnarray}\label{funct_u}
u_{n m}(\rho,\phi,z) = C_{n m}
\frac{w_0}{w(z)}\exp\left[-\tilde{\rho}^2\right]\exp\left[\!\frac{ik\rho^2
z}{2(z^2+z_R^2)}\right]\left(\sqrt{2}\,\tilde{\rho}\right)^{\vert
m\vert}\!L^{\vert m\vert}_n(2\tilde{\rho}^2)\,e^{im\phi}e^{-i(2n+\vert
m\vert+1)\arctan(z/z_R)},
\end{eqnarray}
\end{widetext} where $C_{n m}$ is the normalization constant,
$w(z)=w_0\sqrt{1+(z/z_R)^2}$ is the $z$-dependent radius of the beam,
$\tilde{\rho}$ is the radial coordinate divided by $w(z)$, $L^m_n$ is the
generalized Laguerre polynomial, and $z_R = \pi w_0/\lambda$ is the Rayleigh
length. The functions $u_{n m}$ describe the beam with the projection of the
orbital angular momentum on the propagation axis defined by $m$. They may be
written in the form
\begin{eqnarray}
 u_{n m}(\rho,\phi,z) = (x \pm iy)^{\vert m\vert}f_{n m}(\rho,z),
\end{eqnarray}
where the upper sign corresponds to the positive values of $m$. This leads to
the following formula
\begin{eqnarray}\label{last}
 &(\partial_x + i\partial_y)(x \pm iy)^{\vert m\vert}
 f_{n m}&\nonumber\\
 &=(x\pm iy)^{\vert m\vert\pm 1}\rho^{\mp 1}
 (\partial\!f_{n m}/\partial\rho + (1\mp 1)f_{n m}/\rho).&
\end{eqnarray}
The velocity (\ref{velocity1}) can be obtained by differentiating the phase of
the function $(\partial_x +i\partial_y)u$ but the expression is quite
cumbersome. However, it is clear from Eq.~(\ref{last}) that the function
$(\partial_x +i\partial_y)u$ for positive and for negative values of $m$
carries $m+1$ units of angular momentum in the $z$ direction. Vortex lines
defined in terms of the RS vector run along the $z$ axis and their vorticity
has the strength $m+1$. At first, these results seem to be in disagreement with
the detailed analysis of angular momentum of Laguerre-Gauss beams by Allen,
Padget, and Babiker given in Ref.~\cite{allen} since they have shown that the
additional unit of angular momentum is to be  added to $m$ or subtracted from
$m$ depending on the (right or left) {\em polarization} of the beam. However,
we have broken this symmetry by considering the RS vector ${\bm F}$ and not its
complex conjugate. This (arbitrary) choice has fixed the (positive) sign of the
polarization. With this proviso, our definition of vortex lines in terms of the
RS vector leads to the same results as the analysis of angular momentum. Each
component $u_{nm}$ has only one vortex line associated with the total angular
momentum. However, superpositions of several $u_{nm}$ components, depending on
their composition, may have additional vortex lines.\

The presence of vortex lines in Laguerre-Gaussian beams is due to the definite
angular momentum in the direction of propagation. The same vortex lines appear
also in electromagnetic multipole fields. In this case the RS vector can be
written in the form \cite{zib}
\begin{eqnarray}\label{multipole}
 {\bm F}(x,y,z,t) = e^{-i\omega t}(k + \nabla\times)j_J(k r)({\bm
 r}\times\nabla)Y_{JM}(\hat{\bm r}).
\end{eqnarray}
For the dipole field ($J = 1,\;M = 1$)
\begin{widetext}
\begin{eqnarray}\label{dipole}
 \left({\bm F}(x,y,z,t)\right)^2 = e^{-2i\omega t}(x+iy)^2
 \frac{\left(3 + 4k^2r^2 -2k^4r^4
 + (2k^2r^2 - 3)\cos(2kr) - 6kr\sin(2kr)\right)}{2k^6r^8}.
\end{eqnarray}
\end{widetext}
Thus, the dipole field for $M=1$ exhibits one vortex line along the $z$-axis
(the direction of the angular momentum quantization) with unit vorticity.
Higher multipoles will exhibit vortex lines carrying more units of vorticity,
depending on the value of the $z$ component of the angular momentum.

\section{Conclusions}

The study presented in this paper fully unifies the description of vortex lines
in electromagnetism and in Schr\"odinger wave mechanics. In both cases there is
a single complex function of space and time whose phase generically has
singularities along one-dimensional curves in three-dimensional space --- the
vortex lines. The velocity four-vector $u_\mu$ associated with the phase of the
electromagnetic field plays the same role as the velocity ${\bm v}$ of the
probability fluid in wave mechanics. The circulation around each vortex line is
quantized in units of $2\pi$. There are two important differences. First, the
gradient of the electromagnetic phase does not have any obvious dynamical
interpretation. Second, the electromagnetic field does not vanish identically
on vortex lines but only the two relativistic invariants vanish and the
energy-momentum becomes locally a null four-vector.

Finally, we would like to mention that in principle one should be able to
construct a hydrodynamic form of electrodynamics, analogous to the Madelung
formulation of wave mechanics. The set of hydrodynamic variables for the
electromagnetic field would comprise the components of the energy-momentum
tensor (only five of them are independent, cf., for example \cite{ibb2}) and
the velocity vector $u_\mu$ that carries the information about the phase of the
RS vector. The quantization condition (\ref{circle}) effectively reduces the
information contained in $u_\mu$ to just one scalar function giving finally six
independent functions. However, we have not found a {\em simple} set of
equations for these hydrodynamic-like variables that would be equivalent to
Maxwell theory.

\acknowledgments{We would like to thank Mark Dennis for very fruitful comments
and for making his PhD Thesis available to us. This research was supported by
the KBN Grant 5PO3B 14920.}


\end{document}